\documentclass[aps,twocolumn,prl,showpacs,amsmath,amssymb,superscriptaddress]{revtex4}

\usepackage{graphicx}
\usepackage{dcolumn}
\usepackage{bm}
\usepackage{longtable}
\usepackage{color}
\usepackage{soul}

\makeatletter
\newcommand\figcaption{\def\@captype{figure}\caption}
\newcommand\tabcaption{\def\@captype{table}\caption}
\makeatother

\begin{document}


\title{Profiling the interface electron gas of LaAlO$_3$/SrTiO$_3$
heterostructures by hard X-ray photoelectron spectroscopy}

\author{M. Sing}
\author{G. Berner}
\author{K. Go\ss}
\author{A. M{\"u}ller}
\author{A. Ruff}
\author{A. Wetscherek}

\affiliation{Experimentelle Physik 4, Universit\"at W\"urzburg, Am 
Hubland, D-97074 W\"urzburg, Germany}

\author{S. Thiel}
\author{J. Mannhart}
\affiliation{Institute of Physics, Universit\"at Augsburg, Electronic
Correlations and Magnetism, Experimentalphysik VI, Universit\"atsstrasse 1,
D-86135 Augsburg, Germany}

\author{S. A. Pauli}
\author{C. W. Schneider}
\author{P. R. Willmott}
\affiliation{Paul Scherrer Institut, CH-5232 Villigen, Switzerland}

\author{M. Gorgoi}
\author{F. Sch\"afers}
\affiliation{Berliner Elektronenspeicherring-Gesellschaft f\"ur 
Synchrotronstrahlung m.b.H., Albert-Einstein-Str.~15, D-12489 
Berlin, Germany}

\author{R. Claessen}
\affiliation{Experimentelle Physik 4, Universit\"at W\"urzburg, Am 
Hubland, D-97074 W\"urzburg, Germany}

\date{\today}

\begin{abstract}
The conducting interface of LaAlO$_3$/SrTiO$_3$ 
heterostructures has been studied by hard X-ray photoelectron spectroscopy. From the Ti~2$p$ signal and its angle-dependence
we derive that the thickness of the electron gas is much 
smaller than the probing depth of 4\,nm and that the carrier
densities vary with increasing number of LaAlO$_3$ overlayers. 
Our results point to an electronic reconstruction in the LaAlO$_3$ overlayer as
the driving mechanism for the conducting interface and corroborate the recent interpretation of the superconducting ground state
as being of the Berezinskii-Kosterlitz-Thouless type.
\end{abstract}

\pacs{73.20.-r, 73.40.-c, 73.50.Pz, 79.60.Jv}
\maketitle

Novel phases with often unexpected electronic and magnetic 
properties may form at the interfaces of epitaxial heterostructures 
made out of complex insulating oxides. A case in point is LaAlO$_3$ 
(LAO) on TiO$_2$-terminated SrTiO$_3$ (STO), for which a metallic 
interface state was found at room temperature (RT) \cite{Ohtomo04}. While 
two-dimensional (2D) superconducting behavior below 200\,mK 
\cite{Reyren07,Caviglia08} and indications of ferromagnetism below $\sim 1$\,K 
\cite{Brinkman07} have been reported, the origin and nature of the 
metallic state have been matter of intense debate 
\cite{Ohtomo04,Nakagawa06,Huijben06,Pentcheva06,Thiel06,Brinkman07,Herranz07,
Siemons07,Reyren07,Vonk07,Yoshimatsu08,Basletic08}. 
The metallic state could be either of extrinsic origin, i.e., due to 
effective n-doping by oxygen vacancies, or be intrinsic, i.e., owing 
to electronic reconstruction. In the latter case, the polar 
structure of LAO with alternating (LaO)$^+$ and (AlO$_2$)$^{-}$ 
planes leads to a monotonically increasing potential with increasing 
number of monolayers \cite{Nakagawa06}. To avoid this polar 
catastrophe, in the most simple picture half an electron charge
per 2D unit cell is transferred to the interface \cite{Noguera00}, corresponding
to a sheet 
carrier density $n_{2D}\approx 3.4 \times 10^{14}$\,cm$^{-2}$. However,
depending on sample growing conditions, in 
particular the oxygen partial pressure during deposition, 
experimentally determined sheet carrier densities of $\sim 
10^{13}-10^{17}$\,cm$^{-2}$ 
\cite{Thiel06,Brinkman07,Siemons07,Kalabukhov07,Basletic08} have 
been measured. Likewise, considerable efforts have been made to 
determine the thickness of the conducting layer. Values between 
7\,nm and 600\,${\mu}$m were found 
\cite{Nakagawa06,Herranz07,Siemons07,Reyren07,Willmott07,Basletic08}.  
Unfortunately, simultaneous information on layer thickness and 
charge carrier concentration is scarce 
\cite{Siemons07,Reyren07,Basletic08}.

In this Letter we present for the first time direct spectroscopic 
evidence for the two-dimensional electron gas (2DEG), using hard 
X-ray photoelectron spectroscopy (HAXPES). From the Ti$^{3+}$ signal in the 2$p$ core-level 
spectra we have clear indication for extra electrons located at 
Ti sites. We are further able to extract 
information on layer thickness and carrier density on an equal 
footing. It is found that the surplus interface charge is confined 
to a thickness considerably smaller than the photoemission probing 
depth, probably to one unit cell only. The charge density per 
unit area increases strongly with the number 
of LAO overlayer unit cells. We also observe the creation of extrinsic charge 
carriers by X-ray exposure.
Our results are in favor of electronic reconstruction in the LAO
overlayer instead of band-bending as the driving cause for the 2DEG
formation. The very small 2DEG thickness confirms the generically 2D
nature of the transition into a superconducting ground state which has recently
been proposed for identically prepared samples
\cite{Reyren07}.
\begin{figure}
\begin{minipage}{.23\textwidth}
\centering
\includegraphics[width = 3.9cm]{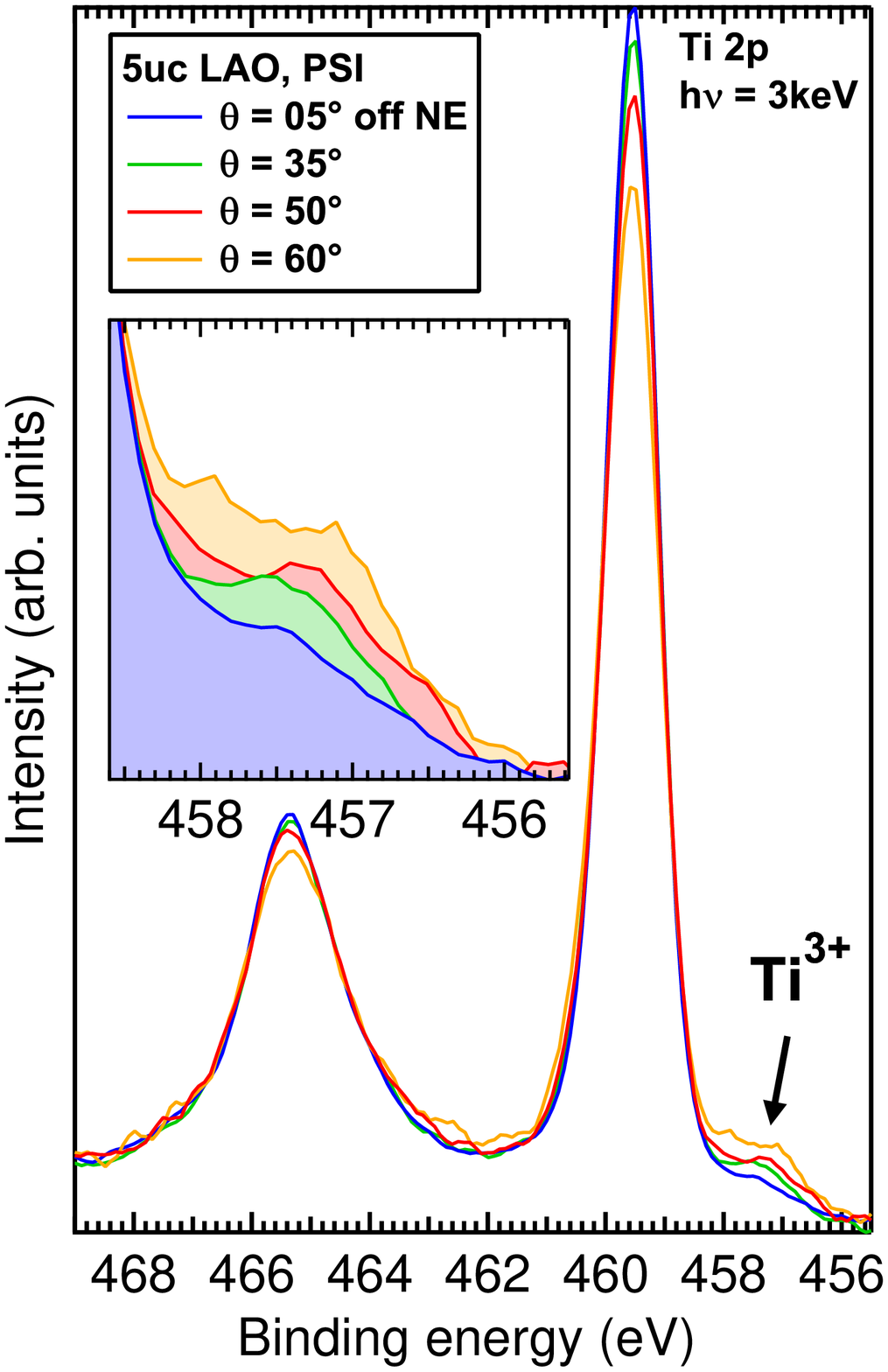}
\end{minipage}
\hfill
\begin{minipage}{.23\textwidth}
\centering
\includegraphics[width = 3.9cm]{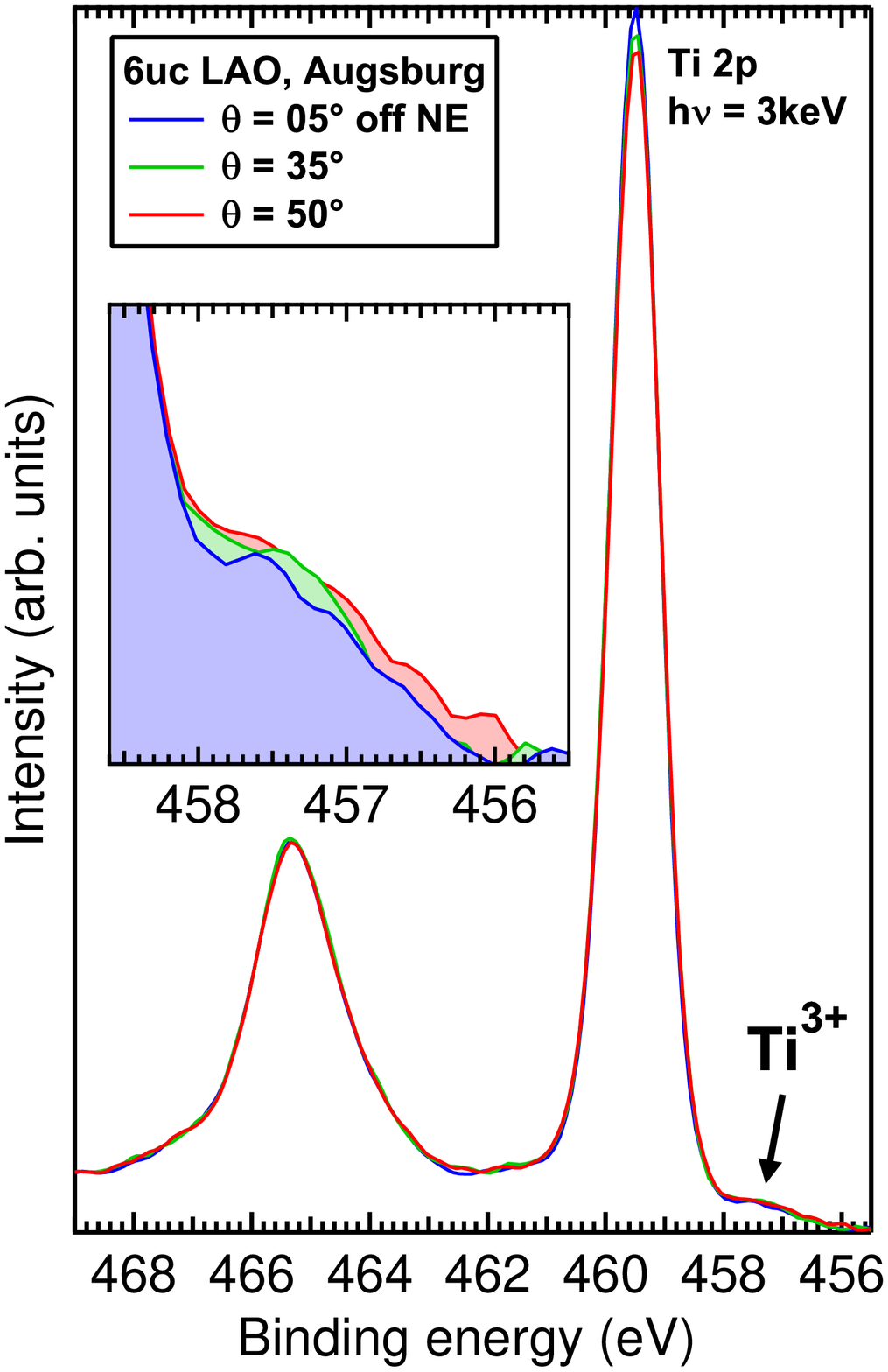}
\end{minipage}
\caption{\label{Figure1} (Color online) Ti~2$p$ spectra of two 
different LAO/STO samples for various emission 
angles $\theta$.}
\end{figure}

LAO/STO heterostructures have been grown by pulsed laser deposition (PLD) on
TiO$_2$ terminated 
(001) STO surfaces. One set of samples 
--- with 2, 4, 5, 6 unit cell (uc) thick LAO overlayers --- was 
grown at the University of Augsburg as described elsewhere 
\cite{Thiel06}. Here we only point out that after deposition in an 
O$_2$ atmosphere of $2\times 10^{-5}$\,mbar around 800$^\circ$C 
these samples were cooled to RT in 400\,mbar of O$_2$ 
with an extra one hour oxidation step at 600$^\circ$C (these 
samples are denoted "Augsburg samples" throughout this paper) in order to 
avoid oxygen vacancies. The other sample is 5\,uc thick and was prepared at the
Paul Scherrer Institute under $5\times 
10^{-6}$\,mbar of O$_2$ and otherwise similar conditions. 
This sample was not subject to a particular oxidation treatment (it is
henceforth denoted "PSI sample") and is the same as that
used in 
Ref.~\onlinecite{Willmott07}.

Hard X-ray photoemission was performed at beamline KMC-1 
\cite{Schaefers07} of the synchrotron BESSY using the endstation HIKE. The total energy 
resolution using 3\,keV photons amounted to $\approx 500$\,meV. 
Binding energies were calibrated with reference to the Au~4$f$ core-level at
$84.0$\,eV. Due to the large probing depth no particular surface preparation was
necessary. The 
Augsburg 4\,uc sample has been contacted as described in 
Ref.~\onlinecite{Thiel06} and allowed 
for {\it in situ} conductivity measurements. All
data was recorded at RT and is
normalized to the background intensity at higher binding 
energies or, equivalently, to equal integrated intensity.

\begin{figure*}
\begin{minipage}[c]{0.25\textwidth}
\centering
\includegraphics[width=0.6\textwidth]{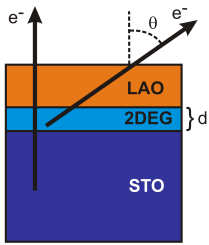}
\end{minipage}%
\hfill
\begin{minipage}[c]{0.7\textwidth}
\begin{ruledtabular}
\begin{tabular}{lccccc}
sample & 2\,uc & 4\,uc & 5\,uc & 6\,uc & 5\,uc \\
& Augsburg & Augsburg & Augsburg & Augsburg & PSI \\
\hline
$p$ (best fit) & 0.01 & 0.05 & 0.02 & 0.02 & 0.28 \\
$p$ range & 0.01\ldots 0.02  & 0.02\ldots0.06 & 0.02\ldots0.09 & 0.02\ldots0.11
& 0.1\ldots0.28 \\
$d$ (uc) (best fit) & $3$ & $1$ & $6$ & $8$ &
$1$
\\
$d$ range (uc) & 1\ldots3  & 1\ldots4 & 1\ldots8 & 1\ldots10 & 1\ldots3 \\
$n_{2D}$ (10$^{13}$\,cm$^{-2}$) & 2.1  & 3.9 & 8.1 & 11.1 & 20.0 \\
\end{tabular}
\end{ruledtabular}
\end{minipage}
\begin{minipage}[t]{0.25\textwidth}
\caption{\label{Figure2} (Color online) Schematic illustrating depth 
profiling by angle-dependent HAXPES.} 
\end{minipage}
\hfill
\begin{minipage}[t]{0.7\textwidth}
\tabcaption{\label{table1} Parameters characterizing the LAO/STO 
interface electron gases from the analysis of the HAXPES data. For 
details see text.}
\end{minipage}
\end{figure*}

In Fig.~\ref{Figure1}, HAXPES spectra are presented of the Ti~2$p$ 
doublet at different emission
angles $\theta$ with respect to the surface normal (normal emission -- NE). The
data sets were 
recorded on PSI (left panel) and Augsburg (right panel) samples 
exhibiting an interface 2DEG. The 
low spectral weight at the lower binding energy side of the main line, detailed
in the insets of Fig.~\ref{Figure1}, can be attributed to emission from
the 2$p$ level of Ti$^{3+}$ as 
evidenced by its energetic shift of 2.2\,eV. Thus it represents a 
direct manifestation of additional electrons hosted in the otherwise 
empty 3$d$ shell of Ti$^{4+}$ in STO. We note that this has not been seen 
before with soft X-ray PES due to the insufficient probing depth 
\cite{Yoshimatsu08}. Going 
to larger emission angles 
--- which corresponds to a decrease in the effective
electron escape depth as $\lambda_{eff}=\lambda \cos \theta$ (see 
Fig.~\ref{Figure2}) --- the Ti$^{3+}$ signal increases in relation 
to the Ti$^{4+}$ main line. From these observations we deduce that 
the extra electrons are localized at the STO side of the 
LAO/STO interface within a region considerably smaller than the
electron escape depth.

For a more quantitative analysis we use the following 
simple model (cf. Fig.~\ref{Figure2}): The 2DEG extends from the 
interface to a depth $d$ into the STO substrate. The
interface 
region is stoichiometric and characterized by a constant fraction $p$ of
Ti$^{3+}$ ions 
per unit cell. Taking into account the exponential damping factor
e$^{-z/\lambda_{eff}}$
for photoelectrons created in depth $z$ one can easily calculate the
ratio of Ti$^{3+}$ to Ti$^{4+}$ signal as a function of emission 
angle $\theta$ (note that the damping in the LAO overlayer does not 
change this ratio anymore but only results in an absolute reduction 
of the signal):
\begin{equation}
\frac{I(3+)}{I(4+)}=\frac{p[1-\exp(-d/\lambda\cos\theta)]}{
1-p[1-\exp(-d/\lambda\cos\theta)]}
\label{eqn1}
\end{equation}
For $d\gg\lambda$ Eq.~\ref{eqn1} reduces to $I(3+)/I(4+)=p/(1-p)$, 
which means that there is no angular dependence in this case. Note
that in Eq.~\ref{eqn1} $p$ and $d$ are not independent. However, 
due to the exponentials $d$ reacts very sensitive 
to a small variation of $p$ except in the limit $d\ll\lambda$, implying
that the parameter range for $p$ and $d$ can be 
narrowed effectively by comparison with experiment.
\begin{figure}
\includegraphics[width = 7.9cm]{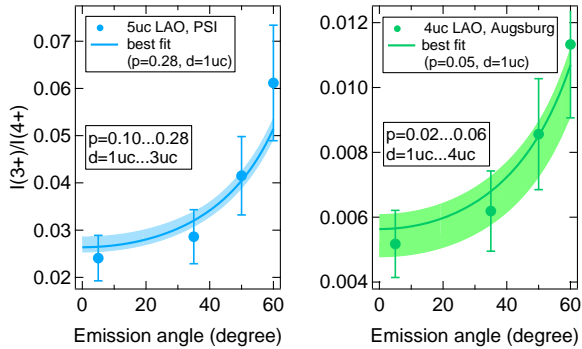}
\caption{\label{Figure3} (Color online) Experimental I(3+)/I(4+) 
ratios for two LAO/STO samples as a function of angle.}
\end{figure}

This is illustrated in Fig.~\ref{Figure3}, where we show the angle 
dependence of the $I(3+)/I(4+)$ ratio for several LAO/STO samples, as 
obtained by a standard fitting procedure. The shaded areas 
mark the array of curves according to Eq.~\ref{eqn1} falling 
within the error bars ($\pm 20$\,\%) of the experimental I(3+)/I(4+) 
ratios. The corresponding parameter ranges for $p$ and $d$ are indicated in
Fig.~\ref{Figure3} and listed in Table~\ref{table1} for all samples. Also
drawn are best fit curves (solid lines). The electron escape depth
$\lambda$ in STO
was fixed to 40\,{\AA} according to the NIST database \cite{NIST} and 
experimental findings on other insulating oxide compounds
\cite{Dallera04,Sacchi05,Wadati08}. As can be seen from the 
parameter ranges compatible with the data, the 2DEG thickness amounts to 
only a few STO unit cells. The carrier concentration is far below 
the expected 0.5\,$e^-$ per unit cell derived from simple 
electrostatics. From the best fit curves there is even a clear trend 
discernible of the 2DEG being confined to only 1\,uc.

\begin{figure}
\includegraphics[width=7.9cm]{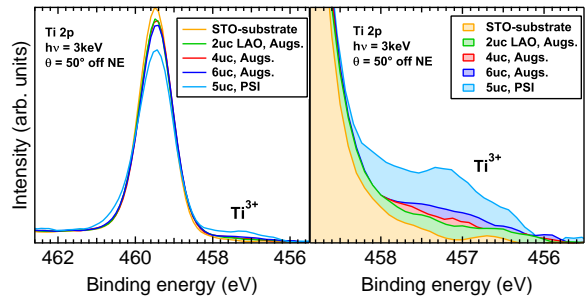}
\caption{\label{Figure5} (Color online) Ti~2$p$ spectra of various samples plus bare STO at fixed emission angle.}
\end{figure}

We now turn to a qualitative comparison of the different samples 
based on their HAXPES spectra. Figure~\ref{Figure5} displays 
Ti~2$p_{3/2}$ spectra (left panel) and their Ti$^{3+}$-related part 
(right panel) at an emission 
angle of 50$^\circ$. For a bare STO substrate there is no sign
of Ti$^{3+}$-related
spectral weight, while there is small but finite weight 
discernible for all other samples. The Ti$^{3+}$ 
intensity steadily increases with the number of LAO overlayers for 
the Augsburg samples and has a maximum for the PSI 5\,uc 
sample.

It is noteworthy that the 2\,uc sample exhibits finite 
charge carrier concentration, although from transport it is
insulating \cite{Thiel06}. The 
finite charge density is in line with {\it in situ} conductivity measurements on
a 4\,uc sample which shows a sharp 
increase by roughly a factor of two upon X-ray exposure. After 
switching the X-rays off, the conductivity relaxes with a time 
constant of several hours. Interestingly, the bare 
STO substrate does not show any sizeable Ti$^{3+}$ spectral weight 
indicating that the LAO/STO interface is important to collect the 
mobile amount of photogenerated electrons. It was argued that in the 
polar discontinuity model the electric potential which accumulates across 
the LAO overlayer must first reach a 
critical value before the activation energy for LAO electrons to 
move to the interface is overcome. This argument offers an 
explanation why highly mobile photogenerated electrons are confined to the
interface already below the
critical LAO thickness of 4\,uc while they are freely distributed and not seen
in HAXPES in the case of bare SrTiO$_3$ with no buried interface. We remark
that part of the
finite charge
carrier concentration in the insulating 2\,uc sample may be of an intrinsic
origin, but not lead to conduction. From the 2DEG thickness ranges in
Table~\ref{table1} a tendency towards increasing thickness with increasing
charge carrier concentration can be inferred. It is well conceivable that a
minimal amount of carriers is needed for
coherent conduction. Indeed, in recent density-functional calculations of the
interface electronic structure two types of charge carriers were found
\cite{Popovic08}. The first is hosted by a 2D band, confined to one interface
layer and hence particularly susceptible to localization by disorder or electron
phonon-coupling, while the other occupies Bloch states, delocalized over several
interface layers, and will contribute to transport.

To extract total charge carrier concentrations $n_{2D}$ one has to 
divide $p\cdot d$ as obtained from the best fit curves by $a^2$, where $a$ is
the STO lattice 
constant. The resulting values for $n_{2D}$ are summarized in 
Table~\ref{table1}. The difference between the 5\,uc Augsburg and 
PSI samples by a factor of 2.5 is conspicuous. Due to the 
lower O$_2$ partial pressure during growth and the absence of an 
extra oxidation step one might invoke extrinsic doping by oxygen 
vacancies as an explanation. However, in this case one would expect 
a much more extended electron gas owing to their high diffusion 
rates \cite{Ishigaki88}. Rather, the subtle 
stoichiometric changes across the interface region previously found 
on this sample \cite{Willmott07} may be responsible for the high 
charge carrier density.

In the Augsburg samples $n_{2D}$ increases with the number of 
LAO overlayers, at variance with Hall measurements. A 
jump at a critical thickness of 4\,uc is observed followed by 
a plateau reaching up to 15\,uc \cite{Thiel06}. This discrepancy could be due to
the fact that the
photogenerated charge carrier contribution is not constant, since with
increasing
LAO thickness and concomitant increasing 
electric potential more electrons photoexcited to trap states are 
released. On the other hand, a further increase of the charge carrier
concentration
does not necessarily have to
result in a higher conductivity if the additionally occupied states are
localized, as outlined above.

The thickness of the 
conducting layer and the charge carrier concentration
are key quantities in gaining information on
both the mechanism responsible for the formation of the 2DEG as well as the
nature of the
possible ground states. We note that the existence of a critical
LAO thickness regarding metallic conductivity precludes any scenario
for the 2DEG formation in the Augsburg samples which is solely based on
extrinsic mechanisms like oxygen deficiencies, interface mixing, or interface
off-stoichiometry. Recently, based on
core-level shifts seen in soft x-ray PES for varying numbers of LAO 
overlayers it was argued that conventional band-bending induces
the mobile charge carriers \cite{Yoshimatsu08}. The fact that
no signal from the metallic states at the chemical potential could be detected
was used to derive a lower limit for the 2DEG
thickness of 5\,nm, at variance with our results. Using qualitative
textbook arguments one expects a
decrease of the space charge region with decreasing dielectric
constant and increasing charge carrier concentration. The dielectric constant
of SrTiO$_3$ strongly decreases with increasing
electric field \cite{Berg95}. From this and the charge carrier
densities given in
Table~\ref{table1}, in a band-bending scenario, the interface
thickness should decrease with the number of LAO overlayers, wheras the
opposite trend is observed (Table~\ref{table1}). Hence, we exclude band-bending
as the driving force for the formation of the 2DEG in our samples.

A scenario in which electronic
reconstruction neutralizes the polar
catastrophe matches our data better. In
this picture, the interface electron gas can be confined to a region as thin
as a unit cell, because the corresponding charge compensation can take
place over the entire LAO overlayer, which acts as a series of parallel-plate
capacitors \cite{Noguera00}. A charge transfer that is smaller than
$0.5e^-$ per 2D unit
cell on the outer ``plates'' can be explained, e.g., by polar lattice
distortions, which screen the local electric field and reduce the band
discontinuity \cite{Nakagawa06,Pentcheva08}, or by
surface adsorbates, which on samples exposed to air are always present. Here we
emphasize that we measure the samples without any
surface preparation and in the lateral center
of the surface. The latter aspect distinguishes
our results from determinations of the 2DEG thickness by spatially resolved
methods \cite{Nakagawa06,Basletic08} which probe the side faces after
mechanical treatment and may be
affected by fringe field
effects. In such techniques significantly larger thicknesses of the 2DEG
($\sim 7$\,nm) are
observed.

For the ground state both a
magnetic \cite{Brinkman07} and a superconducting \cite{Reyren07} phase have been
reported. While in one study it was
argued that the logarithmic temperature dependence of the sheet resistance
might be indicative of scattering of free charge carriers at magnetic
centers (Kondo effect), the other study reported at $\approx 200$\,mK a
resistance drop by
several orders of magnitude. The transition could consistently be explained as a
Berezinskii-Kosterlitz-Thouless crossover-transition into a superconducting
state, a scenario generic for 2D systems, in which long-range ordering is
forbidden
\cite{Mermin66}.
A strict upper limit for the thickness of the conducting layer
of $\cong 10$\,nm was inferred. Our data on essentially the same samples
strongly
support this conclusion. On the other hand, the Kondo scenario
is not linked to low-dimensionality. One thus might conjecture that
either ground state is possible,
depending on the 2DEG thickness as compared to the coherence
lengths for superconductivity and magnetic order, respectively
\cite{Caviglia08}.

In summary, we depth-profiled the interface electron gas of LAO/STO 
heterostructures by means of angle-dependent HAXPES. We find that 
the 2DEG indeed is confined to one or at most a few STO unit cells. 
The extracted carrier sheet densities are in fair agreement with 
previous Hall measurements. Our
data support electronic reconstruction as the driving force for the 2DEG
formation
and is consistent with recent reasonings on the nature of the superconducting
ground state.

\begin{acknowledgments}
We gratefully acknowledge helpful discussions with T. Kopp and J.-M. Triscone
and financial support
by BMBF (05 KS7WW3), DFG (SFB 484), EC (nanoxide).
\end{acknowledgments}


\end{document}